\documentclass{appolb}
\usepackage{epsfig}
\usepackage{dcolumn}
\usepackage{bm}
\usepackage{hyperref} 
\usepackage{multirow} 
\newcommand{\req}[1]{Eq.\,(\ref{eq:#1})}
\newcommand{\labeq}[1]{\label{eq:#1}}
\long\def\symbolfootnotemark[#1]{\begingroup%
\def\thefootnote{\fnsymbol{footnote}}\footnotemark[#1]\endgroup} 
\begin{document}
\title{%
Statistical Hadronization of Multistrange Particles\thanks{
Presented at 50th Cracow School of Theoretical Physics, 
Zakopane by Michal Petr{\' a}{\v n}
}}
\author{%
Michal Petr{\' a}{\v n$^{1,2}$,
Jean Letessier$^{1,3}$,\\
Vojt{\v e}ch Petr{\' a}{\v c}ek$^2$,
Jan Rafelski$^1$}%
\address{%
$^1$Department of Physics, University of Arizona, Tucson, Arizona 85721\\
$^2$Czech Technical University in Prague,\\Faculty of Nuclear Sciences and Physical Engineering\\
$^3$Laboratoire de Physique Th{\' e}orique et Hautes Energies,\\
Universit{\' e} Paris 6 et 7, Paris 75005, France
}
}
\maketitle

\begin{abstract}
We study multistrange hadrons produced in NA49 and STAR 
experiments   at center of mass energies varying 
from $\sqrt{s_{\small{NN}}}=7.61$ GeV to $200$ GeV. We show that the yields of 
$\Xi$, $\overline{\Xi}$ and $\phi$ can help to constrain  the physical conditions present
in the hot dense fireball source of these multistrange hadrons created in  heavy ion collision.
We  address the question of chemical equilibrium of 
individual quark flavors before and after hadronization and offer a few  predictions for  LHC. 
\end{abstract}
  
\section{Introduction}
The Statistical Hadronization Model (SHM) successfully describes 
particle production in heavy ion collisions for a wide range of reaction energies 
studied in the past decade at SPS and RHIC, and has been  widely accepted as a tool 
in the analysis of physical phenomena.
Special features of multistrange particle yields  observed in these experiments inspired this work.

Particular particle ratios are sensitive to the valence quark content of the particles
under consideration.  For this reason  
we can explore the hot dense fireball flavor content. We gain insight
regarding the mechanisms of QGP breakup by considering how 
simple assumptions about dynamics of the hadronization alter the result. 
We extend here our discussion of how the yields of multistrange 
hadrons can be used in such an analysis~\cite{Petran:2009dc}.

We begin with a summary of experimental results considered in section \ref{sec:res} and introduce
the statistical hadronization models in section \ref{sec:SHM}.
Our main interest here will be the ratio of $\Xi(ssq)$ with $\phi(\bar s s)$, 
which we address in detail in subsection \ref{sec:Xiphi}. 
This experimental data will help us establish that there exists a common set of hadronization 
conditions for quite different  reaction conditions, such as centrality, 
and collision energy. 
Based on the universality of SPS and RHIC results we propose in section \ref{sec:prediction} several
 predictions for LHC energies.

\begin{figure}
\begin{center}
\centerline{\hspace*{-0.3cm}
\epsfig{file=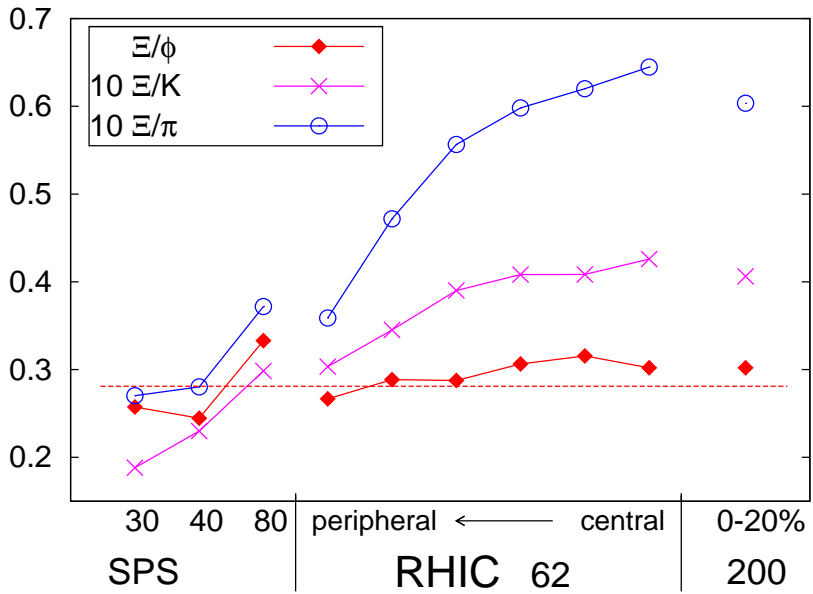,width=3.35in,angle=-90}}
\caption{\label{fig:ratios}  
Data points of  ${\Xi/\phi}$ \req{1}, ${\Xi/{\rm K}}$ \req{5} and  $\Xi/\pi$ \req{8}. 
The straight dashed line shows average ${\Xi/\phi}=0.281$.}
\end{center}
\end{figure}

\begin{table*}[bth]
\caption{\label{tab:table1}Values for the ratios of 
${\Xi/\phi}$ \req{1}, ${\Xi/{\rm K}}$ \req{5},  $\Xi/\pi$ and $\phi/\pi$ \req{8} 
obtained from the experimental data of individual particle yields and the resulting estimated 
uncertainty in $\gamma_q$ and $\gamma_s$ respectively. Symbol `E'  in the error column signals that
the particle ratio is the result of the interpolation and/or extrapolation needed 
to account for different centrality bins.}
\begin{tabular}{lcccccc}
\hline
\vspace{-0.4cm}\\
Experiment 	& Ref. &	Centrality	& 	$10\,{\Xi/\phi}$ 	& 	$\delta\gamma_q$  	& 	$10^2{\Xi/{\rm K}}$ 	&	$\delta\gamma_s$\\
\hline	
STAR 62       & \cite{Abelev:2008zk,Speltz:2006,Abelev:2008ez} &	0-5\%	&	3.04	&	E	&	4.19	&	9.6\%	\\	
STAR 62	& \cite{Abelev:2008zk,Speltz:2006,Abelev:2008ez} &	5-10\%	&	3.00	&	E	&	4.08	&	9.2\%	\\	
STAR 62	& \cite{Abelev:2008zk,Speltz:2006,Abelev:2008ez} &	10-20\%	&	2.94	&	E	&	4.06	&	9.3\%	\\	
STAR 62	& \cite{Abelev:2008zk,Speltz:2006,Abelev:2008ez} &	20-40\%	&	2.88	&	12.5\%	&	3.79	&	E	\\	
STAR 62	& \cite{Abelev:2008zk,Speltz:2006,Abelev:2008ez} &	40-60\%	&	2.85	&	14.6\%	&	3.38	&	E	\\	
STAR 62	& \cite{Abelev:2008zk,Speltz:2006,Abelev:2008ez} &	60-80\%	&	2.49	&	19.3\%	&	2.84	&	E	\\	\hline
STAR 200	& \cite{Abelev:2008zk,Adams:2006ke,Abelev:2008ez} &	0-20\%	&	3.02	&	11.8\%	&	4.06	&	12.9\%	\\	\hline
SPS 80A	& \cite{Alt:2008iv,Alt:2008qm,Afanasiev:2002mx} &	7\%	&	3.33	&	24.5\%	&	3.04	&	22.7\%	\\	
SPS 40A	& \cite{Alt:2008iv,Alt:2008qm,Afanasiev:2002mx} &	7\%	&	2.45	&	42.1\%	&	1.89	&	18.0\%	\\	
SPS 30A	& \cite{Alt:2008iv,Alt:2008qm,:2007fe} &	7\%	&	2.57	&	66.5\%	&	1.85	&	24.3\%	\\ \hline \hline
\vspace{-0.4cm}\\
Experiment 	& Ref. & 	Centrality	&	$10^3\Xi/\pi$	 & 	$10\,\phi/{\rm K}$ 	&	 $10^2\phi/\pi$	&	\\	\hline	
STAR 62	& \cite{Abelev:2008zk,Speltz:2006,Abelev:2008ez} &	0-5\%	&	6.22	&	1.38	&	2.04	&	\\		
STAR 62	& \cite{Abelev:2008zk,Speltz:2006,Abelev:2008ez} &	5-10\%	&	6.20	&	1.36	&	2.06	&	\\		
STAR 62	& \cite{Abelev:2008zk,Speltz:2006,Abelev:2008ez} &	10-20\%	&	5.98	&	1.38	&	2.04	&	\\		
STAR 62	& \cite{Abelev:2008zk,Speltz:2006,Abelev:2008ez} &	20-40\%	&	5.48	&	1.32	&	1.91	&	\\		
STAR 62	& \cite{Abelev:2008zk,Speltz:2006,Abelev:2008ez} &	40-60\%	&	4.65	&	1.18	&	1.63	&	\\		
STAR 62	& \cite{Abelev:2008zk,Speltz:2006,Abelev:2008ez} &	60-80\%	&	3.45	&	1.14	&	1.38	&	\\	\hline	
STAR 200	& \cite{Abelev:2008zk,Adams:2006ke,Abelev:2008ez} &	0-20\%	&	6.04	&	1.34	&	{\symbolfootnotemark[3]$2.54^{+0.21}_{-0.09}$}	&	\\	\hline	
SPS 80A	& \cite{Alt:2008iv,Alt:2008qm,Afanasiev:2002mx} &	7\%	&	2.6	&	0.83	&	0.88	&	\\		
SPS 40A	& \cite{Alt:2008iv,Alt:2008qm,Afanasiev:2002mx} &	7\%	&	3.23	&	0.78	&	0.83	&	\\		
SPS 30A	& \cite{Alt:2008iv,Alt:2008qm,:2007fe} &	7\%	&	2.1	&	0.63	&	0.72	&	\\ \hline
\end{tabular} 
\symbolfootnotemark[3]{
For STAR 200 $\phi/\pi$ considering  figure 14 in~\cite{Abelev:2008zk} we give an average  
of data  for centrality up to 50\%.}
\end{table*}

\section{Experimental results}\label{sec:res}
Data of interest for us obtained in Pb-Pb collisions by NA49 collaboration 
at SPS and in Au-Au collisions by STAR collaboration at RHIC 
are summarized in table \ref{tab:table1}. Using these results we obtained 
the generalized ratios of $\Xi/\phi$ (see \req{1}), $\Xi/K$ (see \req{5}) and $\Xi/\pi$
 (see \req{8})  shown in figure~\ref{fig:ratios}.
We note that the ratio of
$\Xi/\phi$ is practically  constant for all  systems studied
 with center of mass energies varying from $\sqrt{s_{NN}}= 7.61$ to $200$ GeV 
and number of participants ranging from 
$\mathrm{N_{part}}\simeq   20$ (most peripheral collisions) to $ 350$ (most central).

\begin{figure}
\centerline{\hspace*{-0.3cm}
\epsfig{file=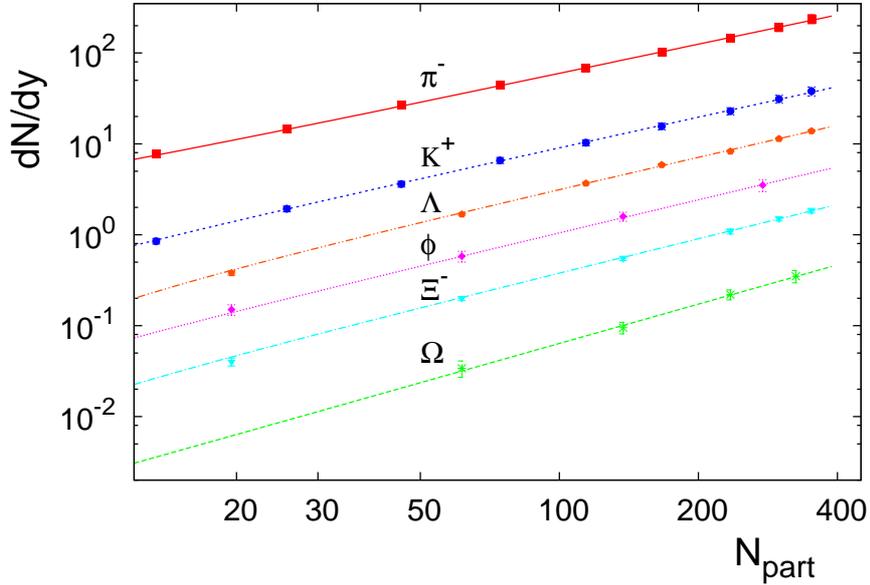,width=3.35in,angle=-90}
}
\caption{\label{fig:fit}  
Data points (full symbols) of particle yields used in the analysis,
and their respective fitted centrality dependence.
}
\end{figure}

We are in particular interested in recent data from RHIC obtained as a function 
of centrality at $\sqrt{s_{NN}}=62.4$ GeV. We performed 
 a centrality dependence study shown in figure~\ref{fig:fit}.
For every particle species the yield can be interpolated by the following form
\begin{equation}
  f(N_{\rm part}) = a \cdot N_{\rm part}^b + c\,.
\labeq{3}
\end{equation}
This function is a good empirical approximation of the individual particle yields. 
From  the functional form of \req{3}
seen in figure~\ref{fig:fit}, one can see that the individual $\Xi$ and $\phi$ yields change by a factor of ten 
for Au-Au collisions at 62 GeV studied in the STAR experiment.
When we look at the higher energy at STAR and the lower energies at NA49, these individual 
yields vary even more.
However, the ratio $\Xi/\phi$ (as defined by \req{1}) remains
within the small range of $0.249 \leq \Xi/\phi \leq 0.304$. 

\section{Statistical hadronization considerations}\label{sec:SHM}
\subsection{Equilibration during hadronization}\label{sec:models}
Within the SHM approach particle yields are obtained by evaluating 
the accessible phase space. The particle density can be written as,
\begin{equation}
\frac{N}{V}=g\frac{4\pi}{(2\pi)^3}T^3 \sum_{n=1}^{\infty} \frac{\gamma^{\pm n}\lambda^n}{n^3}
\left(\frac{nm}{T}\right)^2 {\mathrm K_2}\left(\frac{nm}{T}\right),
\labeq{11}
\end{equation}
where $T$ is the chemical freeze-out temperature, $g$ is the spin degeneracy factor,
$\gamma$ is the phase space occupancy, $\lambda$ is the fugacity factor based on the quark 
content of the given hadron species, and $\mathrm{K_2}$ is the modified Bessel function of the second kind. 

There are several popular models of statistical hadronization. 
They differ by the assumption made about how the 
quark content of the QGP phase is imaged into the valence 
quark content of the final hadron state. 
To begin, one (tacitly) assumes chemical equilibrium
of quarks in the quark--gluon plasma phase. This equilibrium condition
does not have to (and in fact it is argued in the following that it cannot) become chemical equilibrium of the 
emerging hadron gas. 

To allow for varying degree of chemical equilibration 
among final state hadrons one introduces 
the phase space occupancies $\gamma_q$ and $\gamma_s$ which control 
the  overall number of quark anti-quark pairs 
 of the light and strange quarks respectively. The 
fugacity factors $\lambda_q$ and $\lambda_s$ control
the relative chemical equilibrium (i.e., the relative yields of $u,d,s$ 
quarks and their respective anti-quarks). \\
\phantom{ii}(i): In chemical \emph{equilibrium}, slow hadronization is assumed.
During this phase transition, all the flavors of quarks have enough time 
to re-equilibrate into their respective 
hadron gas equilibrium $\gamma_q=1$ and $\gamma_s=1$ ($q = u,d$).
\phantom{i}(ii): \emph{Semi-equilibrium} model takes into account that 
lighter quarks could equilibrate faster in hadronization 
process than strange ones and thus only $\gamma_s\ne 1$, while is fixed to $\gamma_q=1$.  \\
(iii): Chemical \emph{non-equilibrium} assumes a rapid hadronization 
so that not even light quarks have enough time to re-equilibrate. 
Both strange and light quark phase space occupancies are allowed 
to adopt values different from unity as appropriate to describe the observed 
particle yields. Note that when using this most general approach, it is 
possible to deduce the two earlier models as result of data analysis.

\subsection{Multistrange particle production in SHM and  model parameters}\label{sec:parameters}
To be specific, the directly produced yield of $\phi$ can be expressed as
\begin{equation}
\phi\equiv\frac{N_\phi}{V} = 3\frac{4\pi}{(2\pi)^3} T^3 \sum_{n=1}^\infty \frac{\gamma_s^{2n}}{n^3} 
\left(\frac{nm}{T} \right)^2 {\mathrm K}_2 \left(\frac{nm}{T} \right).
\labeq{12}
\end{equation}

It is important to see how fast this series converges. 
It depends not only on the mass to temperature ratio,
but also on the phase space occupancies and fugacities 
as they may be greater than unity and appear in increasing powers in the series.
To get a sense for how fast this sum converges, let us examine the following case (table~\ref{tab:table2}).

\begin{table}[h!bt]
\begin{center}
\caption{\label{tab:table2} Example of parameters used to study particle densities.}      
\begin{tabular}{lcl}

\hline
Volume & V &  1000 $\rm fm^3$ \\
Temperature & T & 140 MeV \\
$u,d$ Phase space occupancy & $\gamma_q$ & 1.6 \\
$s$ Phase space occupancy  & $\gamma_s$ & 2.2 \\
Fugacities  & $\lambda_{q,s}$ &      1 \\
\hline

\end{tabular}
\end{center}
\end{table}

When we calculate the partial contributions to the particle yields of interest,
we find as shown in table \ref{tab:table3}, 
that the heavier the particle is, the faster the sum converges. 
We have to be careful with less massive particles,
such as pions, and even kaons. Considering the first seven terms
for pions  gives us only 96\% of their yield. For Kaons, we have to take 
the first two terms to get 99.8\% of their yield.
However, for particles whose mass $m \gg 100 \rm MeV/c^2$, 
it is sufficient to take the Boltzmann approximation, {\it i.e.\/},
just the first term $n=1$ in \req{11}:
\begin{equation}
\frac{N}{V}=g\gamma\frac{4\pi}{(2\pi)^3} \lambda m^2T\, {\mathrm K_2}\left(\frac{m}{T}\right).
\labeq{6}
\end{equation}
One has to be even more careful when using  
expansion of the Bessel function in $m/T$ which converges much slower.

\begin{table}[hbt]
\caption{\label{tab:table3} Contribution of individual terms in the sum  \req{11}  to 
the particle yields.}
\begin{tabular}{lcccc}
\hline\hline
              & $\Xi^-$     &  $\phi$     &      $K^-$  & $\pi^-$ \\ \hline
$m[\rm{MeV/c^2}$]     & 1321        &  1020       &      492    &      139 \\
$m/T$         & 9.4357      &  7.2857     &   3.5143    &      0.99286 \\
N             & 0.9862      &  5.6488     &   51.4969   &      219.911 \\ \hline
\multicolumn{5}{c}{ Fractions of contributions order by order:} \\
n=1           & 0.9947       &    0.9992  &    0.9692  &      0.6861 \\
n=2           & $1.98\times10^{-4}$ & $1.04\times10^{-3}$ & $2.89\times10^{-2}$ & 0.138 \\
n=3           & $6.46\times10^{-8}$ & $1.81\times10^{-6}$ & $1.52\times10^{-3}$ & $5.74\times10^{-2}$ \\
n=4           &                     &                     &                     & $3.14\times10^{-2}$ \\
...           &                     &                     &                     & \\
n=7           &                     &                     &                     & $9.78\times10^{-3}$ \\ \hline
\end{tabular}
\end{table}


\subsection{$\Xi/\phi$ ratio}\label{sec:Xiphi}
Experimentally measured particle yield consists not only of directly produced
particles, but also a contribution from heavier resonances decaying 
into it and thus increasing its apparent yield.  In general the experimental 
results comprise these contributions. Therefore, we include in our calculations
directly produced $\Xi$'s and $\phi$'s and the contributions from decays of
heavier resonances. The most significant contribution
comes from $\Xi^*(1530)$ with spin degeneracy $g=4$. We use the decay 
tables as implemented in the SHARE program~\cite{Torrieri:2004zz}.

Particle yield ratios are very sensitive to the valence quark content and we 
can take advantage of this by considering specific ratios. One uses
ratios since when one takes the same number 
of hadrons in the numerator as in the denominator,
the overall normalization is canceled. We also consider
the product of a particle and its antiparticle in order to 
eliminate the chemical potentials, which are opposite,  
and therefore the fugacity factors also cancel.
Finally, by the choice of hadrons involved, one can change how the ratio
depends on the phase space occupancies.

For these reasons, we look first at the generalized 
ratio of $\Xi$ to $\phi$, defined as:
\begin{equation} 
   \frac{\Xi}{\phi} \equiv {\sqrt{\frac{\overline{\Xi}^+\Xi^-}{\phi\phi}}}
       \simeq \frac{\gamma_q \gamma_s^2\lambda_s\lambda_s^{-1}\lambda_q^{1/2}\lambda_q^{-1/2}}
       {\gamma_s^2\lambda_s\lambda_s^{-1}} f (T,m_\phi,m_\Xi)
       = \gamma_q  f (T,m_\phi,m_\Xi).
\labeq{1}
\end{equation}
The ratio of $\Xi/\phi$ is proportional to 
the light quark phase space occupancy $\gamma_q$ and a function $f$ of 
temperature $T$, 
\begin{equation}
f (T,m_\phi,m_\Xi) \simeq  \frac{\sum\limits_i g_{\Xi_i} m_{\Xi_i}^2 K_2\left(\frac{m_{\Xi_i}}{T} \right)}
{\sum\limits_j g_{\phi_j} m_{\phi_j}^2 K_2\left(\frac{m_{\phi_j}}{T} \right)},
\labeq{13}
\end{equation}
which in the Boltzmann approximation is determined by degeneracies $g_i$ and 
masses of the hadrons contributing to the yields and over which we sum using the Boltzmann approximation.

As evident from \req{11}, the effect of quantum statistics is 
decreasing with the hadron mass $m$ when compared to the 
freeze-out temperature $T$. We calculated the magnitude of this effect 
to be $0.25\%$ for the $\Xi/\phi$ ratio and is therefore negligible. 
One has to remember that the formula for the residual 
function \req{13}  does not account for strong decays in which flavor content 
is redistributed, nor does reflect any possible weak decays
 influencing the yields in question.

Global fits~\cite{Rafelski:2004dp,Rafelski:2009jr,Letessier:2005qe}
 of the systems under consideration show that all these systems
hadronize at the same temperature $T$ in the non-equilibrium model. 
This, together with the observed constant ratio of $\Xi/\phi$, implies
common hadronization conditions for all the systems studied.
Another immediate implication is that the light quark
phase space occupancy $\gamma_q$ is constant and has 
the same value for all the energies, 
centralities and systems studied.

\begin{figure}
\centerline{\hspace*{-0.3cm}
\epsfig{width=3.35in,angle=-90,file=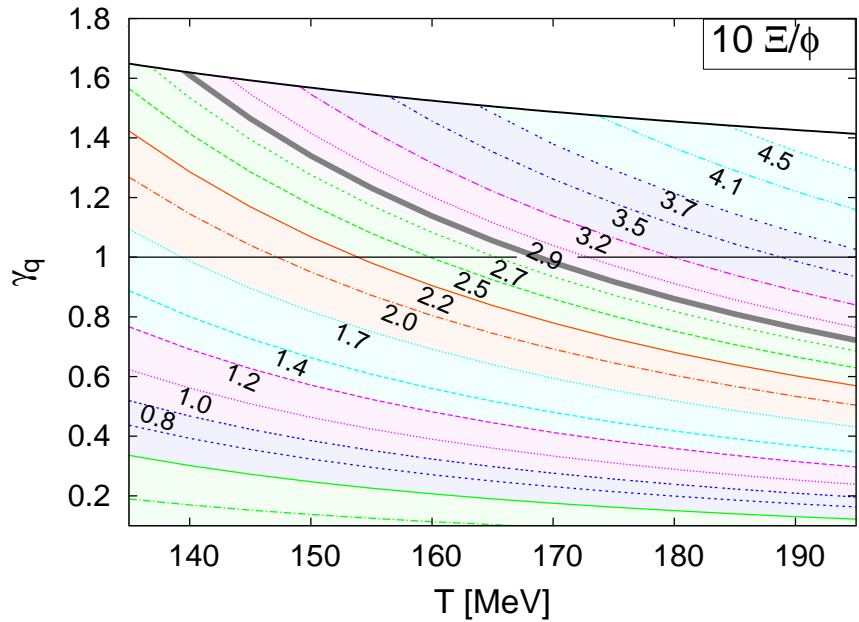}
}
\caption{\label{fig:1}  
 Lines of a constant ratio $10\Xi/\phi \in [0.8,4.5]$ (as defined in \req{1})
in the $T$--$\gamma_q$ plane.
The lines for $\gamma_q = 1$ and $\gamma_q = \gamma_q^{\rm crit}$ 
are presented by solid black lines. 
The thick gray line represents the average result of all SPS and RHIC.}
\end{figure}

To obtain the results presented here we actually use the SHARE program \cite{Torrieri:2004zz}. 
We consider several values of the $\Xi/\phi$ ratio and
plot the results in the $\gamma_q$--$T$ plane in figure \ref{fig:1}. 
In combination with the global fit results, there are two possible regions 
to look for possible hadronization conditions. 
The first region corresponds to light quarks being in chemical equilibrium. 
This is represented 
by $\gamma_q=1$, and from figure~\ref{fig:1}, we see temperature $T\simeq 170$ MeV.
To locate the second region, we allow for fast hadronization 
and relax light quark phase 
space occupancy  $\gamma_q\ne 1$. 
When varying the magnitude of $\gamma_q$, one has to remember
that there is a critical value of this parameter 
for which pions begin to form a Bose--Einstein
condensate. This critical value is given by the mass 
of the lightest, non--strange meson ($\pi^0$) by:
\begin{equation}
 \gamma_q < \gamma_q^{crit} \equiv \mathrm{exp}\left(\frac{m_{\pi^0}}{2T} \right).
\labeq{7}
\end{equation}
This is shown as an upper limit in figure~\ref{fig:1}. 

In conclusion of this analysis, we observe that it is possible for different
reaction systems to occupy different values of $T$ and $\gamma_q$ along the solid black line in 
figure~\ref{fig:1}. However, much more plausible interpretation is that all reaction systems 
considered hadronize at the same set of $T,\gamma_q$. This result is strongly supported by 
the global fit of all hadron yields we performed and in which near critical value of
$\gamma_q \simeq 1.6$ is favored. Figure~\ref{fig:1} then implies temperature $T\simeq 140$ MeV.

\subsection{Hadronization of strangeness}
In the previous subsection, we found that the ratio $\Xi/\phi$ is proportional
to light quark phase space occupancy $\gamma_q$.
To study strangeness occupancy $\gamma_s$ we have to 
identify a ratio proportional to $\gamma_s$.
By the same reasoning as before, we are motivated to examine the ratio 
of $\Xi$ to K defined as:
\begin{equation}
 {\Xi\over{\rm K}} \equiv\! 
{\sqrt{\frac{\overline\Xi^+\Xi^-}{{\rm K}^+{\rm K}^-}}}=\gamma_s f_1(T). 
 \labeq{5}
\end{equation}

It is evident from both table~\ref{tab:table1} and figure~\ref{fig:ratios} that
$\Xi$/K is not constant.  The Au-Au,   $\sqrt{s_{NN}}=62.4$ GeV,   RHIC data which 
inspired this study show a systematic variation with a factor of 1.5 difference 
between central and peripheral collisions. When we consider the other systems
and energies studied,  the factor increases to more than two. We believe this is
due entirely to variation in $\gamma_s$ since as we argued that $T=$ Const.

When $\gamma_s$ is allowed to vary, we have to remember, especially 
in the view of the expected conditions at LHC where $\gamma_s$ could turn out to be large, 
that there is also a critical value of $\gamma_s$. It turns out that it is the $\eta$ meson 
($\eta = 0.45\mathrm{(u\bar u+d\bar d) + 0.55 s\bar{s}}$) which is the first to condensate when 
$\gamma_s$ increases. 
The critical value for $\gamma_s$ is a function of temperature and can be obtained from
the condition:
\begin{equation}
0.45\gamma_q^2 + 0.55\gamma_s^2 = \mathrm{exp}\left( \frac{m_{\eta}}{T}   \right).
\labeq{14}
\end{equation}
The critical value of $\gamma_s$ is not as sensitive to $\gamma_q$ as it is to temperature.
For the two scenarios discussed, $T=140\,\rm{MeV}$ and $\quad T=170\,\rm{MeV}$,  
we get the critical value of $\gamma_s^{crit} = 9.47$ and $\gamma_s^{crit} = 6.68$ respectively.

\section{Predictions}\label{sec:prediction}
\subsection{General considerations}

We have argued in the last section, that the temperature 
is constant for all the systems, and we have attributed the 
change in relative yield of $\Xi$ and K to the change in $\gamma_s$.
This implies that the strangeness initially present in 
the fireball of QGP does not have enough time to re-equilibrate 
during the hadronization, and the strange phase space in the 
hadron gas phase can be under- or over-populated. As a consequence, 
we conclude that chemical equilibrium model (i) in subsection \ref{sec:models},
which assumed both light 
and strange quarks to be in chemical equilibrium after hadronization, is no longer 
a viable description of particle yields in heavy-ion collisions.

With the range of  $\gamma_q, \gamma_s$ and  $T$ established by measuring $\Xi/\phi$ and $\Xi/\rm{K}$,
we can now predict the ratios of particles which are harder to observe experimentally.
According to the quark content of the particles in question, we establish how their ratio 
depends on phase space occupancies and with that predict the behavior and possible values of 
their ratio.

\begin{figure}[h!tb]
\centerline{
\epsfig{file=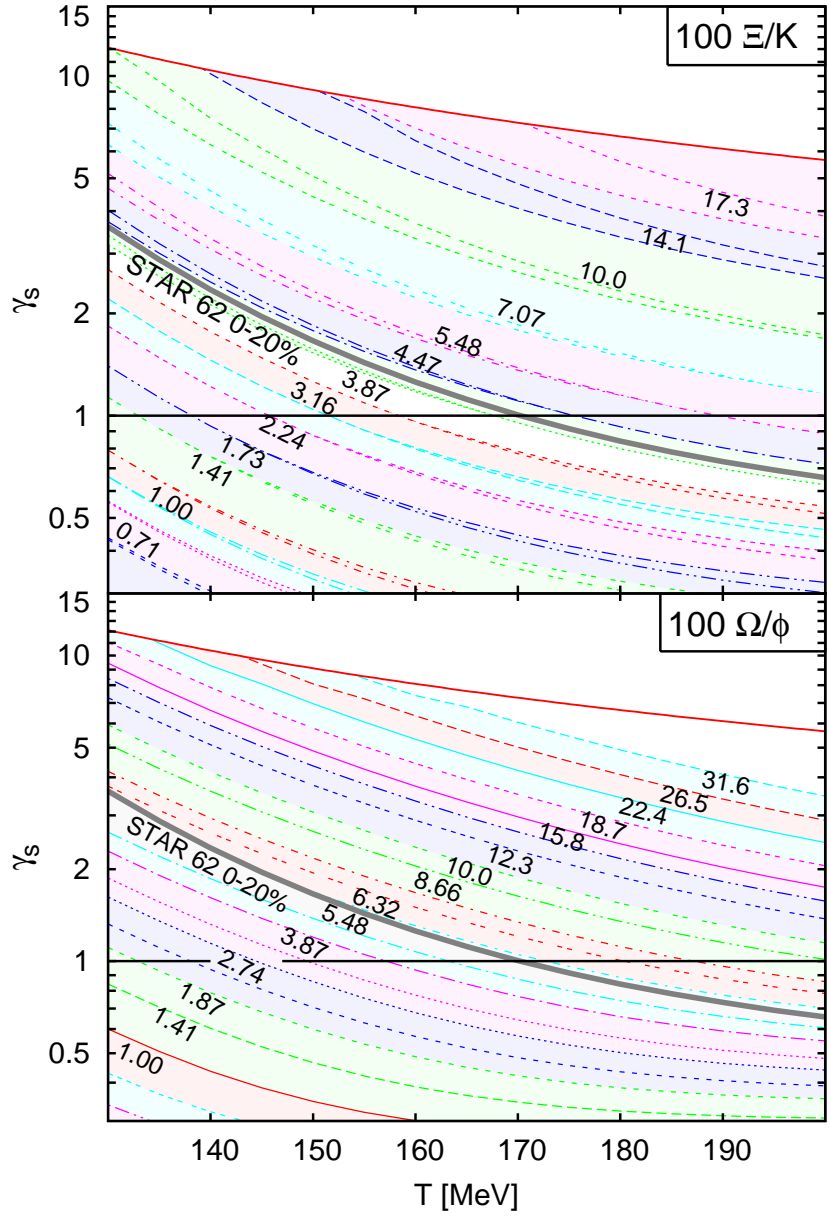,width=6.5in, angle=-90}
}
\caption{\label{fig:3} Lines of constant ratio ${\Xi/{\rm K}}$ (top) 
and ${\Omega/\phi}$ (bottom).
Experimental data from most central  0--20\% STAR 62 are indicated by a thick  line (top), 
and is assumed in the bottom frame as a prediction. See text for more detail.  }
\end{figure}

Both the ratios $\Xi/\mathrm{K}$ and $\Omega/\phi$ are proportional to $\gamma_s$.
In the top section  of figure~\ref{fig:3} we show fixed values of $\Xi/\mathrm{K}$ and 
the band of experimental data for this ratio in the $T-\gamma_s$ plane. 
In the bottom frame of figure~\ref{fig:3} we show the corresponding $\Omega/\phi$ ratio 
and find a narrow range:
\begin{equation}
\frac{\Omega}{\phi}\equiv\!\sqrt{\frac{\Omega^-\overline{\Omega}^+}{\phi\phi}}
 \in \langle 5.5\times 10^{-2}, 7.0\times 10^{-2}\rangle.
\labeq{9}
\end{equation}

Similarly, with temperature $T$, $\gamma_q$ and $\gamma_s$ constrained to a small range of values,
we can predict other ratios of particles, such as $\phi/\pi$  
or $\Xi/\pi$ (see figure \ref{fig:ratios}) defined by
\begin{equation}
\frac{\Xi}{\pi}\equiv \sqrt{\frac{\Xi^-\overline \Xi^+}{\pi^-\pi^+}} \propto \frac{\gamma_s^2}{\gamma_q};\quad
\frac{\phi}{\pi}\equiv \sqrt{\frac{\phi\phi}{\pi^-\pi^+}} \propto \left( \frac{\gamma_s}{\gamma_q} \right)^2,
 \labeq{8}
\end{equation}
which are both very sensitive to the amount of strangeness 
in the emerging hadron gas due to their dependence on $\gamma_s^2$.

\subsection{LHC conditions}
One very simple and solid prediction we can make based on the universality of the 
results for $\Xi/\phi$ is that $\Xi/\phi = 0.28\pm 10$\% at LHC too. 
This universal hadronization condition for $T,\gamma_q$ 
has been observed so far in both SPS and RHIC results. 
We strongly believe that this yield will be observed in heavy ion reactions at LHC. 

We also expect considerably greater absolute yield of $\phi$ at LHC, since one can anticipate 
that there is a greater strange quark pair density production because of more extreme 
initial conditions~\cite{Letessier:2006wn}. We found 
in a kinetic exploration of strangeness production  
that the strangeness over entropy
ratio at LHC is at least $s/S \simeq 0.037$. 
This implies, as shown in~\cite{Kuznetsova:2006bh}, that 
the ratio of $\gamma_s/\gamma_q \simeq 1.55$. We use this value in our further LHC predictions.
Here we note that since the size of 
reaction volume changes, as does baryon and strangeness content, several SHM parameters
such as $V, \gamma_s, \lambda_q, \lambda_s$ are expected to change. Thus the hadronization
universality only refers to $T,\gamma_q$.

\begin{figure}
\centerline{\hspace*{-0.2cm}
\epsfig{file=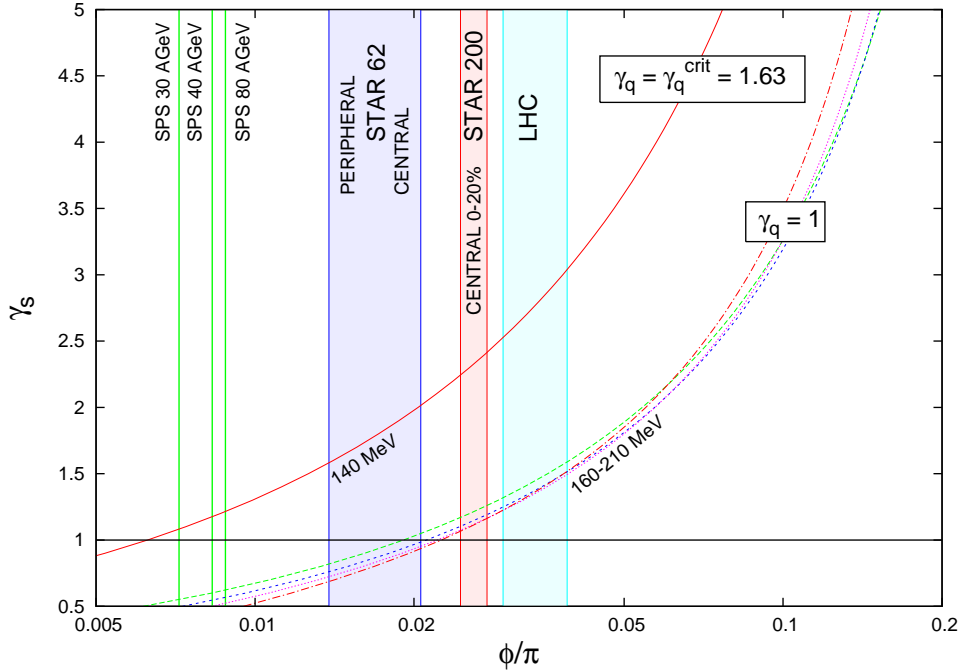,width=3.55in,angle=-90}
}
\caption{\label{fig:4}   
$\gamma_s$ as a function of the relative $\phi/\pi$ yield \req{8} in 
two hadronization scenarios, see text. 
The horizontal solid black line shows the chemical 
equilibrium with $\gamma_s = 1$. For 
experimental data see table \ref{tab:table1}. }
\end{figure}

We show the required value of $\gamma_s$ as a function of the measured 
value of  $\phi/\pi $ in figure~\ref{fig:4} for both 
chemical semi-, and non-equilibrium.  Also we show (colored) bands of data from 
different experiments and also a prediction
for LHC results. It is evident from the figure that the temperature 
does not have a great effect on this ratio when we fix $\gamma_q = 1$, a very fortuitous
result --- the increase of $\phi$-yield with $T$ is nearly compensated by increase in $\pi$-yield.
Thus, if at LHC $\phi/\pi> 0.03 $ we can be sure that irrespective of the model
of hadronization the value of $\gamma_s>1$. 

$\gamma_s>1$ cannot be build up from a small initial value in hadron based
strangeness production since chemical equilibrium is reached at $\gamma_s=1$\,.
On the other hand, when the source of hadrons is a strangeness rich initial phase, such
as chemically equilibrated quark--gluon plasma,  hadronization can lead to phase space
overpopulation in final state. This is true in particular for a fast hadronization process,
in which strangeness content does not have time to re-annihilate.

\section{Conclusions}
The
experimental results from SPS and RHIC provided an important observation, that the ratio
$\Xi/\phi$ is nearly constant for wide range of energies, 
centralities and even across different systems, while 
the individual yields change by more than an order of magnitude. 
As this ratio is governed by  light quark phase space occupancy $\gamma_q$, 
and hadronization temperature $T$, than these conditions are very likely universal 
for  all the  experiments. This universality in regard to $T,\gamma_q$ 
is strongly supported by  global 
data fits~\cite{Letessier:2005qe,Rafelski:2009gu}. 

A different ratio, $\Xi/\mathrm{K}$, which is proportional
 to strange phase space occupancy $\gamma_s$,
changes by a factor of 2, which implies that $\gamma_s$ 
changes significantly between different 
experiments and centralities. Thus, we learn that 
the full chemical equilibrium hadronization is not a viable model.

We further find that some experimentally observed yields of multistrange 
hadrons, such as $\phi$, are only compatible with $\gamma_s>1$. 
For this to happen, we require strangeness rich initial
phase (quark--gluon plasma) and fast hadronization, so that 
strangeness abundance does not have enough time to 
re-equilibrate during hadronization.

Our study allowed us to predict certain 
particle ratios for which precise data is not available at present experiments 
and for the upcoming LHC. For example, we expect 
$\Omega/\phi \in \langle 5.5\times 10^{-2}, 7.0\times 10^{-2}\rangle$ at RHIC 62 GeV.
We expect the ratio $\Xi/\phi$ to remain in the same range of values at LHC as it 
has been  measured at SPS and RHIC,  i.e., $\Xi/\phi\in \langle 0.249,0.304\rangle$.\\[0.8cm]

\noindent Laboratoire de Physique Th{\' e}orique et Hautes Energies, LPTHE, at University Paris
6 and 7 is supported by CNRS as Unit{\' e} Mixte de Recherche, UMR7589. This work was 
supported by the grant LC07048 and LA 316 from Czech Ministry of Education and the grant from
the U.S. Department of Energy, DE-FG02-04ER41318.

\end{document}